%% file: main.tex
\documentclass[]{mosi}

\usepackage{helvet}

\usepackage{amsmath} 
\usepackage{natbib}
\usepackage{graphicx}
\usepackage{subcaption} 

\usepackage[toc,page,header]{appendix}
\usepackage[utf8]{inputenc} 
\usepackage[T1]{fontenc}    
\usepackage{hyperref}       
\usepackage{url}            
\usepackage{booktabs}       
\usepackage{lmodern}        
\usepackage{amsfonts}       
\usepackage{nicefrac}       
\usepackage{microtype}      
\usepackage{wrapfig}

\usepackage{amssymb}        
\usepackage{titletoc}
\usepackage{minitoc}

\usepackage{array}
\usepackage{etoolbox}

\definecolor{lightblue}{RGB}{200, 230, 255}  
\definecolor{headerblue}{RGB}{150, 200, 255} 

\usepackage{pgfplots}
\usepackage{xcolor}
\usepackage{float}
\usepackage{comment}
\usepackage{multirow}
\usepackage{makecell}
\usepackage{siunitx}
\usepackage{tikz}
\usepackage{pgf-pie}
\usepackage[export]{adjustbox}

\usepackage{ragged2e}
\usepackage{tabularx}
\usepackage{caption}
\usepackage{enumitem}
\usepackage{pifont}
\usepackage[hang,flushmargin]{footmisc}

\usepackage{tcolorbox}
\tcbuselibrary{breakable}
\tcbuselibrary{skins}

\usepackage{tabularx}
\usepackage{listings}

\usepackage{threeparttable}
\usepackage{setspace}
\usepackage[scheme=plain]{ctex} 

\definecolor{MossCyan}{HTML}{82D9FF} 
\definecolor{MossBlue}{HTML}{82B1FF}

\definecolor{ForestGreen}{RGB}{34, 139, 34}
\definecolor{Red}{RGB}{255, 0, 0}

\definecolor{tickG}{rgb}{0.1, 0.588, 0.1}
\definecolor{crossR}{rgb}{0.588, 0.1, 0.1}
\newcommand{\cmark}{\textcolor{tickG}{\ding{52}}}

\definecolor{frenchblue}{rgb}{0.0, 0.45, 0.73}
\definecolor{babyblue}{rgb}{0.54, 0.81, 0.94}
\definecolor{classicrose}{rgb}{0.98, 0.8, 0.91}
\definecolor{beige}{rgb}{0.96, 0.96, 0.86}
\definecolor{forestgreen}{HTML}{2e7d43}

\definecolor{blue1}{HTML}{91BBE6}
\definecolor{blue2}{HTML}{3F90E0}
\definecolor{blue3}{HTML}{316FAD}

\definecolor{color1}{HTML}{FF9999}
\definecolor{color2}{HTML}{FF6666}
\definecolor{color3}{HTML}{FF3333}
\definecolor{color4}{HTML}{E60000}
\definecolor{color5}{HTML}{B30000}
\definecolor{color6}{HTML}{8CD98C}
\definecolor{color7}{HTML}{53c653}
\definecolor{color8}{HTML}{00B050}
\definecolor{color9}{HTML}{2d862d}
\definecolor{color10}{HTML}{206020}
\definecolor{color11}{HTML}{cca300}

\usepackage{booktabs}   
\usepackage{colortbl}   
\usepackage{xcolor}     
\usepackage{pifont}     
\usepackage{graphicx}   

\usepackage{booktabs}   
\usepackage{longtable}  
\usepackage{hyperref}   
\definecolor{lightgreen}{RGB}{235, 255, 235} 
\definecolor{checkgreen}{RGB}{0, 150, 0}     
\usepackage{graphicx}
\usepackage{wrapfig}
\usepackage{caption}
\input{txt_display}
\usepackage{needspace}

\newtcolorbox{promptbox}[2][]{
    colback=white,
    coltext=black,
    arc=3mm,
    boxrule=0.5pt,
    colframe=black!60!white,
    title={#2},
    colbacktitle=black,
    coltitle=white,
    fonttitle=\bfseries,
    top=8pt,
    bottom=8pt,
    left=10pt,
    right=10pt,
    breakable,
    before upper={%
        \linespread{1}\selectfont
        \setlength{\parskip}{1ex plus 0.2ex minus 0.2ex}%
        \setlength{\parindent}{0pt}%
    },
    #1
}

\title{ABC-Bench: Benchmarking Agentic Backend Coding\\in Real-World Development}

\author{
    Jie Yang\textsuperscript{1,*} \quad
    Honglin Guo\textsuperscript{1,*} \quad
    Li Ji\textsuperscript{1} \quad
    Jiazheng Zhou\textsuperscript{1} \quad
    \textbf{
    Rui Zheng\textsuperscript{2} \quad
    \\
    Zhikai Lei\textsuperscript{2} \quad
    Shuo Zhang\textsuperscript{2} \quad
    Zhiheng Xi\textsuperscript{1} \quad
    Shichun Liu\textsuperscript{1} \quad
    Yuxin Wang\textsuperscript{1} \quad
    \\
    Bo Wang\textsuperscript{1} \quad
    Yining Zheng\textsuperscript{1,\dag} \quad
    Tao Gui\textsuperscript{1,\dag} \quad
    Xipeng Qiu\textsuperscript{1,3}
    }
    \affiliation{
        \textsuperscript{1}Fudan University \quad
        \textsuperscript{2}Shanghai Qiji Zhifeng Co., Ltd. \quad
        \textsuperscript{3}Shanghai Innovation Institute
    }
}

\abstract{
\begin{abstract}
'The evolution of Large Language Models (LLMs) into autonomous agents has expanded the scope of AI coding from localized code generation to complex, repository-level, and execution-driven problem solving. However, current benchmarks predominantly evaluate code logic in static contexts, neglecting the dynamic, full-process requirements of real-world engineering, particularly in backend development which demands rigorous environment configuration and service deployment. To address this gap, we introduce ABC-Bench, a benchmark explicitly designed to evaluate agentic backend coding within a realistic, executable workflow. Using a scalable automated pipeline, we curated 224 practical tasks spanning 8 languages and 19 frameworks from open-source repositories. Distinct from previous evaluations, ABC-Bench require the agents to manage the entire development lifecycle from repository exploration to instantiating containerized services and pass the external end-to-end API tests. Our extensive evaluation reveals that even state-of-the-art models struggle to deliver reliable performance on these holistic tasks, highlighting a substantial disparity between current model capabilities and the demands of practical backend engineering.
\end{abstract}
}

\checkdata[Github]{\url{https://github.com/OpenMOSS/ABC-Bench}}
\checkdata[Dataset]{\url{https://huggingface.co/datasets/OpenMOSS-Team/ABC-Bench}}

\begin{document}
\maketitle
\begingroup
\renewcommand{\thefootnote}{\fnsymbol{footnote}}
\setcounter{footnote}{1}
\footnotetext{Equal Contribution. Work done during an internship at Shanghai Qiji Zhifeng Co., Ltd.}

\setcounter{footnote}{2}
\footnotetext{Corresponding authors. \url{ynzheng@fudan.edu.cn}, \url{tgui@fudan.edu.cn}}
\endgroup




\section{Introduction}
Recent advancement of Large Language Models (LLMs) has redefined the role of AI in software engineering: moving beyond simple code prediction to acting as autonomous agents capable of exploring repositories, wielding terminal tools, and executing complex tasks within real environments.~\citep{traeagent2024, yang2024sweagent, openhands2024, bouzenia2024repairagent, zhang2024codeagent, lin2025se}. Consequently, there is an urgent need to evaluate their ability in handling real-world software engineering tasks.

\begin{figure*}[t]
  \centering
  \includegraphics[width=1.0\textwidth]{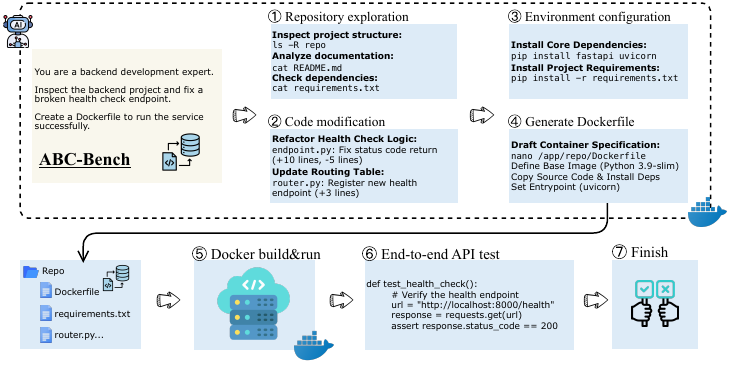}
  \caption{\textbf{Overview of the ABC-Bench evaluation pipeline.} The figure illustrates the closed-loop evaluation process. In the development phase (Steps 1--4), the agent acts as a backend expert to analyze the repository, resolve issues, and draft container specifications. Transitioning to the validation phase, the benchmark system builds a Docker image from the agent's output and deploys the service (Step 5). Finally, the functional correctness is verified by sending real HTTP requests to the deployed endpoint (Step 6), ensuring the fix works in a production-like environment.}
  \vspace{-1.5em}
  \label{fig:evaluation}
\end{figure*}

Despite the progress on software engineering benchmarks, evaluating these agents in production-like settings remains a critical gap. Current benchmarks focus on localized engineering, such as making isolated code edits that often overlook environment configuration, while relying on fragmented, unit-level validation instead of end-to-end tests~\citep{naturalcodebench,zan2025multi, zan2024swe, jin2024swebench, zhou2024feabench, zhang2024pybench,  guo2024codeeditorbench, deng2025swe}. However, real-world software engineering is an integrated workflow where coding, configuration, and deployment are inherently intertwined.\citep{li2024devbench, padigela2025ml, li2025prompting}. This limitation is particularly pronounced in backend development, which demands the integration of code changes with environment configuration and container orchestration. While BaxBench~\citep{vero2025baxbench} attempts to evaluate backend capabilities, it is still constrained to relatively isolated tasks that fail to capture the full complexity of production environments. Consequently, there remains a significant lack of evaluations that rigorously assess backend agents within comprehensive, production-grade workflows.

\input{tables/compare}

To address these needs, we introduce ABC-Bench, a benchmark designed to evaluate agentic backend coding throughout the entire backend development lifecycle. Each task goes beyond localized code edits and requires the agent to configure the environment and instantiate a containerized service. Once the service is launched, we evaluate correctness strictly via external API-level integration tests, awarding credit only when the deployed service starts correctly and exhibits the expected behavior. To construct the benchmark at scale, we build ABC-Pipeline, a task-generation workflow that produces candidate tasks from open-source backend repositories with greatly reduced manual intervention. Applied to 2,000 open-source repositories, it yields 224 curated tasks spanning 8 backend programming languages and 19 frameworks, preserving the heterogeneity of real-world backend stacks. 

Our extensive evaluation of various models and agent frameworks reveals that current systems remain far from reliable on these full-lifecycle tasks. Even the top-performing model achieves only a 63.2\% pass@1 rate, while many others lag substantially, indicating significant headroom for improvement. Further analysis identifies environment configuration and deployment as persistent bottlenecks, often acting as the primary barriers to success. These findings underscore a clear gap between current model capabilities and the rigorous demands of real-world backend engineering.

We highlight our contributions as follows:

\begin{itemize}
    \item We introduce ABC-Bench, a benchmark of 224 full-lifecycle backend tasks requiring autonomous repository exploration, environment configuration, deployment, and API-based verification.
    \item We design ABC-Pipeline, a scalable workflow that automates the extraction of tasks from GitHub, significantly lowering the barrier for constructing realistic evaluation datasets.
    \item We evaluate diverse agents and models, establishing robust baselines and uncovering that environment configuration and deployment are the predominant bottlenecks, offering key insights for future system improvements.
\end{itemize}

\section{Related Work}


\textbf{Agentic Paradigms for Software Engineering.} Recent progress in code-oriented large language models has expanded their scope from isolated programming problems to full-fledged software-engineering tasks. Earlier work on code-focused LLMs primarily emphasized code reasoning and synthesis in competitive or self-contained settings~\citep{qwen25coder,codeneedcomments}, whereas more recent efforts increasingly adopt an agentic paradigm, enabling models to navigate real-world codebases and development environments through tools and multi-step execution to address end-to-end development tasks. A representative line of work conceptualizes software development as an agentic loop in which language models iteratively inspect repositories, modify code, and refine solutions based on execution feedback. This paradigm has been instantiated across a range of systems, from early agent–computer interfaces for resolving real GitHub issues to more general-purpose code-agent frameworks~\citep{yang2024sweagent,livesweagent,openhands2024,traeagent2024}. Collectively, these efforts signal a decisive shift from passive code generation toward autonomous or semi-autonomous software-engineering agents operating over full repositories, moving large language model research beyond proof-of-concept demos toward real-world scenarios with tangible productivity impact, and motivating evaluations grounded in realistic software-development settings.

\textbf{Evaluation of Coding Capability.} In line with the shift toward agentic software-engineering systems that iteratively explore repositories, edit code, and execute tools, evaluation of coding capability has also moved from function-level generation on small, synthetic tasks~\citep{chen2021evaluating,austin2021program} toward more realistic settings that emphasize executable feedback, natural prompts, multi-stage development, and code editing~\citep{rahman2025beyond,yu2024humaneval,mceval,naturalcodebench,li2024devbench,guo2024codeeditorbench}. More recently, repository-level benchmarks evaluate code agents on real-world repositories via multi-step issue resolution and feature implementation under executable test signals~\citep{aleithan2024swe,yu2025utboost,zhang2024pybench}. To rigorously assess agentic mastery of the full software lifecycle, evaluations must extend beyond local code logic to include system-level operations and runtime interactions. Backend development exemplifies these challenges, necessitating an orchestrated workflow from repository exploration to environment provisioning, deployment, and live service validation. However, as shown in Table~\ref{tab:benchmark-coverage}, existing benchmarks typically cover only parts of this lifecycle and rarely evaluate whether agents can deliver a deployment-ready backend service in a realistic stack~\citep{vero2025baxbench}. Consequently, current evaluations often overemphasize code-only changes under pre-configured sandboxed environments and unit-test signals, leaving the full backend agentic workflow underexplored; ABC-Bench closes this gap by assessing the complete backend lifecycle with end-to-end, deployment-oriented verification.

\clearpage
\section{ABC-Bench}
\label{sec:benchmark}

\begin{figure*}[t]
  \centering
  \includegraphics[width=\textwidth]{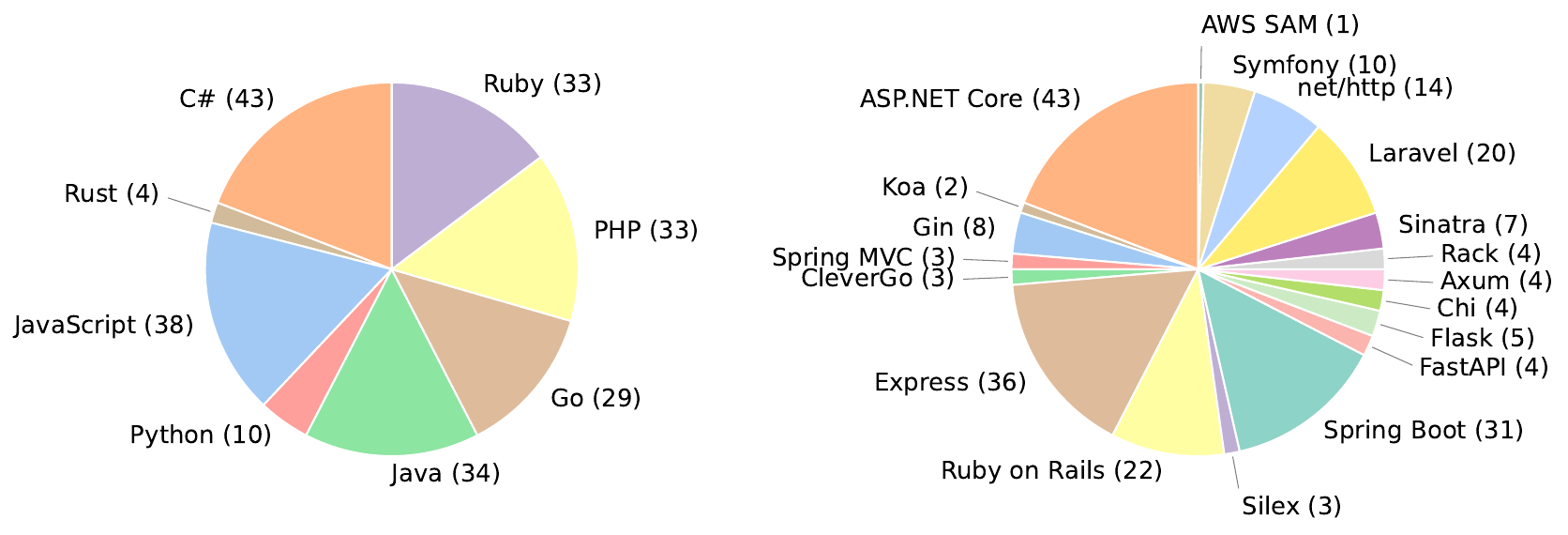}
  \caption{\textbf{Overview of the ABC-Bench dataset composition.} The left pie chart illustrates the distribution of tasks across eight major programming languages. The right chart provides a detailed breakdown of the 19 web frameworks involved, demonstrating the benchmark's capability to evaluate models across a wide spectrum of real-world software stacks.}
  \label{fig:language-and-framework-full}
\end{figure*}

\subsection{Overview of ABC-Bench}
\label{subsec:overview}
ABC-Bench comprises 224 tasks, offering a diverse and balanced representation of modern backend ecosystems. As illustrated in Figure~\ref{fig:language-and-framework-full}, the benchmark spans 8 programming languages and 19 backend frameworks. Of these tasks, 132 focus primarily on logic implementation within a pre-provisioned runtime, while 92 additionally require autonomous environment configuration and containerized service startup, thereby testing end-to-end operational capability.
Beyond technical heterogeneity, the tasks are drawn from a broad spectrum of real-world domains, ensuring that the benchmark reflects practical engineering needs. These domains range from data analytics and search systems to commerce platforms, payment gateways, and developer tooling. Detailed category-level statistics and domain descriptions are summarized in Table~\ref{table:task-category}.

\subsection{Evaluation Pipelines}
\label{sec:eval-pipeline}
We evaluate models and agents using a standardized, isolated sandbox environment, which strictly separates the agent's workspace from the backend service under test. As summarized in Figure~\ref{fig:evaluation}, the evaluation setup launches an outer container that hosts the agent, delivers the task prompt.

Within this workspace, the agent is granted full autonomy to explore the repository, modify code, install dependencies, and update Docker configurations, enabling a comprehensive, full-lifecycle development loop. The agent must orchestrate the entire process without human intervention.

Upon submission of the solution or when the interaction budget limits are reached, the evaluation system attempts to build and launch the backend service in a separate inner container using the agent's generated code and configurations. We determine success exclusively via functional API-level verification by executing external requests against the deployed service to validate its behavior. This execution-driven approach ensures that agents are evaluated not on static code artifacts, but on their ability to deliver a functioning, production-ready service.

\section{ABC-Pipeline}
\label{sec:pipeline}

\begin{figure*}[t]
  \centering
  \includegraphics[width=1.0\textwidth]{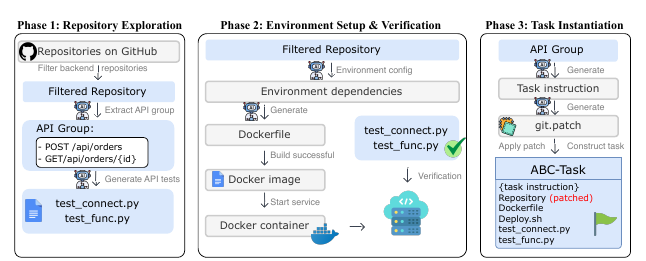}  \caption{\textbf{Overview of the ABC-Pipeline workflow.} The process consists of three phases: (1) Repository Exploration, where backend repositories are filtered and API tests are generated; (2) Environment Setup \& Verification, which involves synthesizing Docker environments and verifying them against generated tests; and (3) Task Instantiation, where the final benchmark task is constructed by applying git patches and generating task instructions.}
  \label{fig:pipeline}
\end{figure*}

To construct realistic backend development tasks at scale, we build the ABC-Pipeline, an automated workflow that converts open-source backend repositories into full-lifecycle development Tasks, as shown in Figure~\ref{fig:pipeline}.

\subsection{Task Construction}
\label{sec:task_construction}

The ABC-Pipeline constructs backend tasks from real-world repositories by leveraging an autonomous agent that augments GPT-5 with command-line capabilities. This construction process is structured into three distinct phases:

\textbf{Phase 1: Repository Exploration.}
We initiate the process by filtering a pool of 2,000 open-source, MIT-licensed repositories to isolate high-quality backend candidates. The construction agent then autonomously explores each repository to identify functional API groups. Rather than relying on existing tests—which may be incomplete or outdated—the agent proactively generates dedicated verification suites. These suites cover both service connectivity and functional logic, serving as the criteria to verify the correctness of the model's solution in the final evaluation pipeline.

\textbf{Phase 2: Environment Synthesis.}
Once API targets are identified, the pipeline advances to runtime synthesis. The agent analyzes the repository structure to resolve dependencies and generates the necessary container configuration files. It then attempts to build the runtime image and launch the service within an isolated container. This phase focuses strictly on establishing a deployable infrastructure, ensuring that the service can start up and listen on the expected ports, preparing the stage for subsequent validation.

\textbf{Phase 3: Task Instantiation.}
In the final phase, the pipeline synthesizes the actual benchmark problems using a masking-based strategy. For a selected API group, the agent formulates a natural language task instruction and synthesizes a solution patch representing the correct implementation. We then apply a reverse operation, selectively masking the implementation logic of the target endpoint to simulate a pre-implementation state. The resulting ABC-Task package encapsulates this masked repository, the task instructions, the environment setup files, and the verification suite. Specifically for tasks designated as environment configuration challenges, we subsequently remove all synthesized environment setup files from the final package and supplement the task instructions with explicit requirements for the evaluated model to autonomously configure the runtime environment.

\subsection{Task Verification}
\label{sec:task_verify}
To guarantee the reliability and solvability of the automatically constructed tasks, we implement a rigorous two-stage verification protocol.

First, we verify the correctness of the ground truth environment and the test suite. We deploy the service using the original, unmasked repository and execute the generated tests. Since the implementation logic is fully intact, a valid task requires the tests to pass successfully. This step serves as a quality gate, filtering out repositories with unstable runtime configurations or flawed test logic, ensuring that the baseline evaluation setup is sound.

Second, we verify the effectiveness of the generated patch (mask) and the test suite's ability to detect missing logic. We apply the masking strategy to the repository to simulate the pre-implementation state, redeploy the service, and re-execute the tests. A valid task requires the tests to fail upon applying the mask. Instances that continue to pass are discarded, as this indicates that either the mask failed to remove the core functionality, or the tests are insufficient to detect the incomplete implementation.

\subsection{Task Distribution}
\label{sec:task_distribution}
Through this pipeline, we initially generated 600 candidate tasks. To ensure a balanced evaluation benchmark, we filtered these candidates based on the distribution of programming languages and frameworks, resulting in a final set of 224 high-quality tasks spanning 8 programming languages and 19 web frameworks, of which 92 additionally require autonomous environment configuration.

\section{Experiments}
\label{sec:experiments}
\subsection{Experimental Setup}
\label{subsec:experimental-setup}
We evaluate a diverse suite of representative open-source and proprietary models. The open-source models include Qwen3-8B and Qwen3-32B~\citep{qwen3_2025}, DeepSeek-V3.2~\citep{deepseekv32_2025}, and GLM 4.7~\citep{glm47_2025}, alongside coding-specialized variants like Qwen3-Coder-30B-A3B-Instruct and Qwen3-Coder-480B-A35B-Instruct~\citep{qwen25coder,qwen3_2025} and agent-oriented models like Nex-N1-671B and Nex-N1-32B~\citep{nexn1_2025}. For proprietary models, we include GPT-5, Gemini 2.5 Pro~\citep{gemini25_2025}, and Claude Sonnet 4.5. For detailed information about the models used in our experiments, please refer to Table~\ref{table:model-detail}.

To ensure a unified evaluation protocol, we employ OpenHands~\citep{openhands2024} as the default agent framework for all models. For each model-agent pairing, we perform three independent runs per task. We set the sampling temperature to $0.7$ for standard models and $1.0$ for reasoning-enhanced variants.

\subsection{Main Results}
\label{subsec:main-results}
\input{tables/main-results}

\textbf{Full-Lifecycle Tasks Remain Challenging.}
Table~\ref{table:main-results} reveals that ABC-Bench presents a rigorous barrier for current models. The state-of-the-art Claude Sonnet 4.5 achieves an overall pass@1 of $63.2\%$, while other models like DeepSeek-V3.2 hover around $50\%$. In contrast, smaller models, such as Qwen3-8B, fail to reach $10\%$. This performance stratification underscores the complexity of full-lifecycle software engineering: unlike isolated code generation, agents must maintain consistency across environment setup, dependency management, and functional deployment.

\textbf{Models Lack Multilingual Robustness.} Performance varies significantly by language stack. While widely-used languages like Python, Go, and JavaScript generally see higher success rates, other languages create major bottlenecks. Rust stands out as an extreme case where most models—including strong open-source contenders like DeepSeek-V3.2 fail to solve a single task (scoring $0.0\%$). Only the most capable proprietary models, specifically Claude Sonnet 4.5 and GPT-5, achieve meaningful success (above $30\%$), highlighting a distinct capability gap in handling less common or more complex language stacks.

\begin{figure*}[t]
  \centering
  \begin{minipage}[t]{0.48\textwidth} 
    \vspace{0pt} 
    \centering
    \includegraphics[width=\linewidth]{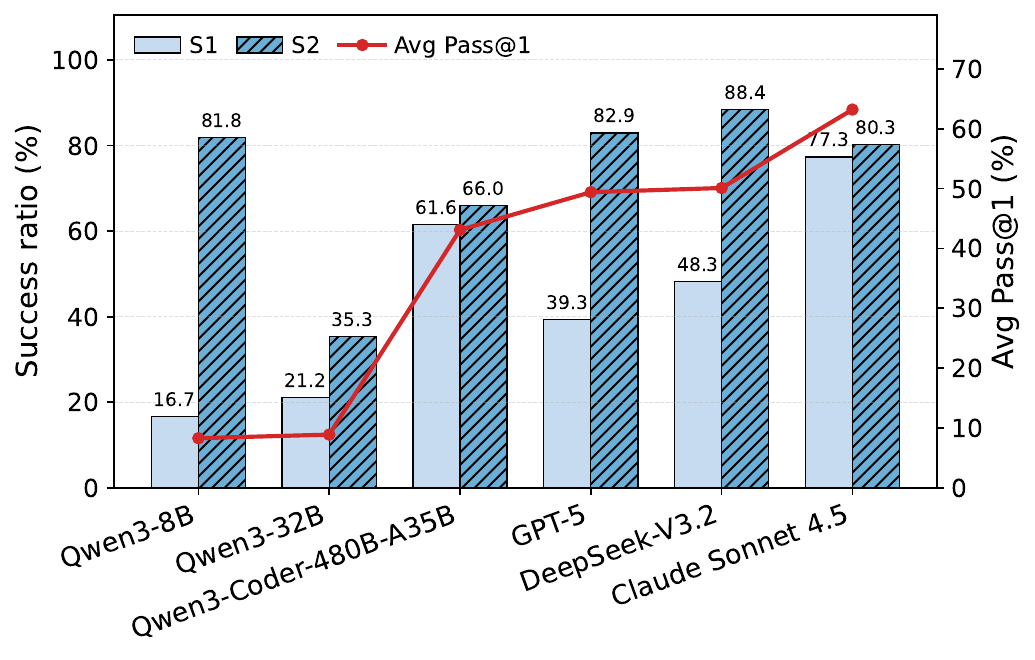}
    \caption{
      \textbf{Analysis of environment configuration capabilities.} 
      Comparison of various models (including Claude Sonnet 4.5, GPT-5, DeepSeek-V3.2, and Qwen3-8B) across 92 environment-related tasks. 
      The bar charts display Build Success ($S_1$) and Conditional End-to-End Success ($S_2$). The red line plot indicates the Average Pass@1.
    }
    \label{fig:env-analysis}
  \end{minipage}
  \hfill
  \begin{minipage}[t]{0.48\textwidth} 
    \vspace{0pt} 
    \centering
    \includegraphics[width=\linewidth]{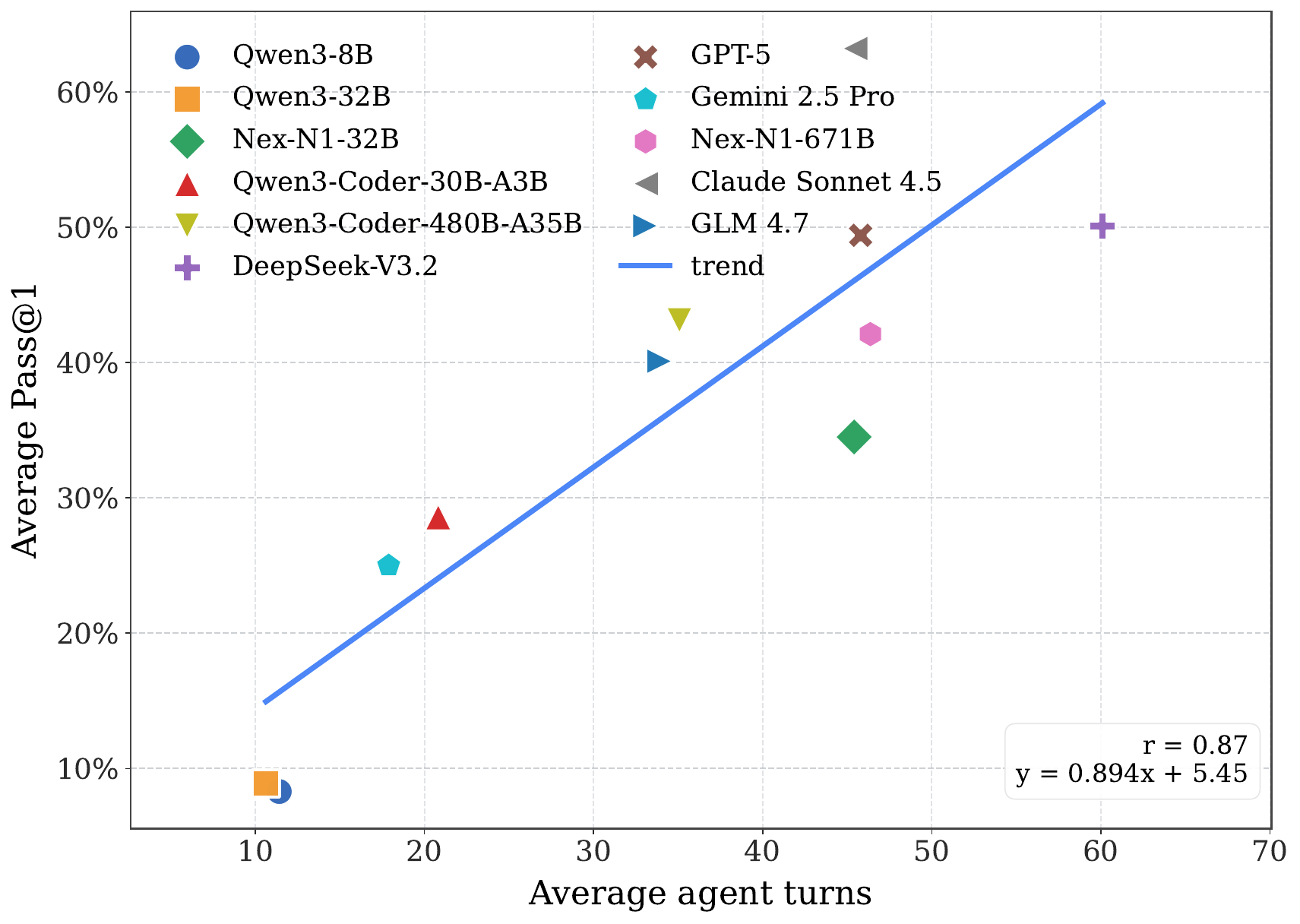}
    \caption{
      \textbf{Interaction turns vs.\ performance.} 
      Scatter plot illustrating the relationship between the average number of agent turns (x-axis) and the Average Pass@1 rate (y-axis) across various models. The blue fitted trend line reveals a strong \textbf{positive correlation} ($r=0.87$).
    }
    \label{fig:turns_vs_perf}
  \end{minipage}
\end{figure*}

\textbf{Environment Configuration as the Primary Bottleneck.} To understand the performance disparity, we analyzed 92 environment-related tasks by decomposing the workflow into two distinct stages: Environment Build ($S_1$), which verifies if the service can be successfully constructed and started, and Functional Execution ($S_2$), which measures the pass rate of functional tests specifically for the subset of tasks that successfully passed $S_1$. As illustrated in Figure~\ref{fig:env-analysis}, Claude Sonnet 4.5 demonstrates the most robust full-lifecycle capability across these tasks, achieving high success rates in both stages ($S_1 \approx 78\%$, $S_2 \approx 80\%$). In contrast, models like GPT-5 and DeepSeek-V3.2 exhibit a striking imbalance: while they excel at functional coding ($S_2 > 80\%$), they struggle significantly with the initial setup ($S_1 < 50\%$). This sharp drop reveals that environment configuration is the primary bottleneck masking their algorithmic proficiency. Smaller models like Qwen3-8B are largely filtered out at the first stage ($S_1 < 20\%$). These findings suggest that bridging the gap in environment configuration is crucial for unlocking the full potential of LLMs in end-to-end software development.

\textbf{Correlation Between Interaction Depth and Success.} In addition to environment capabilities, we observe a strong positive correlation ($r=0.87$) between the depth of agent interaction and task success, as shown in Figure~\ref{fig:turns_vs_perf}. Top-performing models, notably Claude Sonnet 4.5, exhibit the longest execution trajectories (averaging $>60$ turns), whereas weaker models like Qwen3-8B tend to terminate prematurely ($\approx 10$ turns). This trend underscores the iterative nature of ABC-Bench: unlike simple Q\&A tasks, full-lifecycle software engineering requires agents to actively explore the environment, interpret error logs, and perform multiple rounds of debugging. Consequently, the ability to sustain and manage long-horizon interactions acts as a prerequisite for the high performance observed in our main results.

\section{Analysis}
\label{sec:analysis}

\subsection{Agent Frameworks}
\label{subsec:agent-frameworks}

To assess how agent framework influences performance beyond the model itself, we benchmark DeepSeek-V3.2 and GPT-5 across three frameworks: OpenHands, Claude Code, and mini-SWE-agent. Figure~\ref{fig:agent-frameworks-and-sft-effect}(a) reveals that the framework is a critical variable. OpenHands facilitates peak performance ($\approx 50\%$) for both models, whereas mini-SWE-agent causes severe degradation, notably dropping GPT-5's success rate to below $20\%$. This highlights that ABC-Bench evaluates the holistic system, where the framework's interaction strategy is as vital as the underlying model.

\Needspace{8\baselineskip}
\subsection{Effect of Agentic Post-Training}
\label{subsec:sft-effect}

\begin{wrapfigure}{r}{0.5\textwidth}
  \vspace{-30pt}
  \centering
  \includegraphics[width=0.50\textwidth]{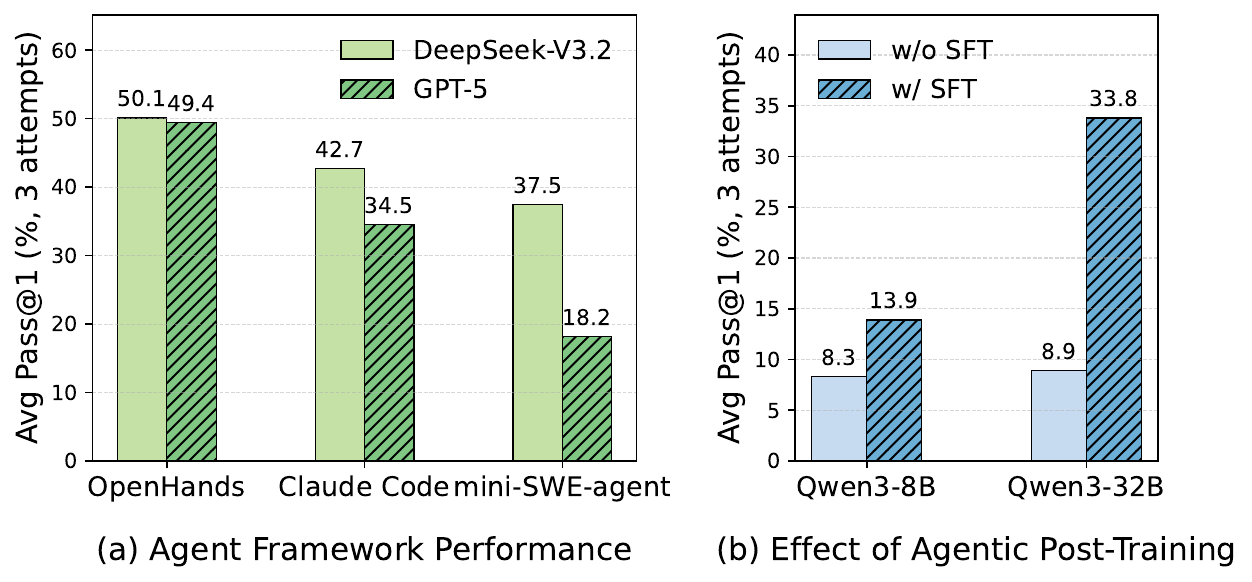}
  \caption{\textbf{(Left) Agent Framework Performance.} Comparison of agent frameworks on ABC-Bench for DeepSeek-V3.2 and GPT-5. \textbf{(Right) Effect of Agentic Post-Training.} Impact of agent-style supervised fine-tuning (SFT) on ABC-Bench. Both are reported as average pass@1 (\%) over three attempts per task.}
  \label{fig:agent-frameworks-and-sft-effect}
  \vspace{-20pt}
\end{wrapfigure}

We investigate the impact of further agentic supervised fine-tuning (SFT) on ABC-Bench performance.
Using the agentic coding subset of the publicly-available \footnote{https://huggingface.co/datasets/nex-agi/agent-sft}{Nex-N1 dataset}, we fine-tuned Qwen3-8B and Qwen3-32B and evaluated them using OpenHands.
Figure~\ref{fig:agent-frameworks-and-sft-effect}(b) demonstrates substantial improvements: pass@1 increases from $8.3\%$ to $13.9\%$ for the 8B model, and surges from $8.9\%$ to $33.8\%$ for the 32B model.
These results indicate that training on high-quality agentic trajectories significantly boosts the model's capability to handle full-lifecycle tasks, with larger models showing particularly strong data efficiency.

\subsection{Performance by Task Category}
\label{subsec:category-effects}

To investigate whether models possess generalized engineering capabilities or exhibit domain-specific biases, we analyze performance variations across different task categories. Figure~\ref{fig:category-heatmap} stratifies performance across diverse application domains, revealing significant variance in task difficulty and model competency. We observe a distinct difficulty hierarchy: models generally excel in Analytics (where Claude Sonnet 4.5 reaches $86.7\%$) and Specialized tasks. In contrast, DevTools—the largest category ($N=53$)—proves consistently challenging, with even the strongest model scoring below $50\%$ ($47.8\%$). This suggests that tasks involving development tooling and infrastructure require more complex, context-heavy reasoning than standard analytical scripts.

Furthermore, the heatmap highlights domain-specific specialization. While Claude Sonnet 4.5 dominates most categories, it is notably outperformed by GPT-5 in the Identity domain ($60.0\%$ vs. $73.3\%$). Overall, these results confirm that ABC-Bench captures meaningful domain heterogeneity, requiring agents to possess robust expertise across a wide spectrum of backend scenarios.

\subsection{Error Analysis and Failure Modes}
\label{subsec:error-analysis}

\begin{figure*}[t]
  \centering
  \begin{minipage}[t]{0.48\textwidth}
    \vspace{0pt} 
    \centering
    \includegraphics[width=\linewidth]{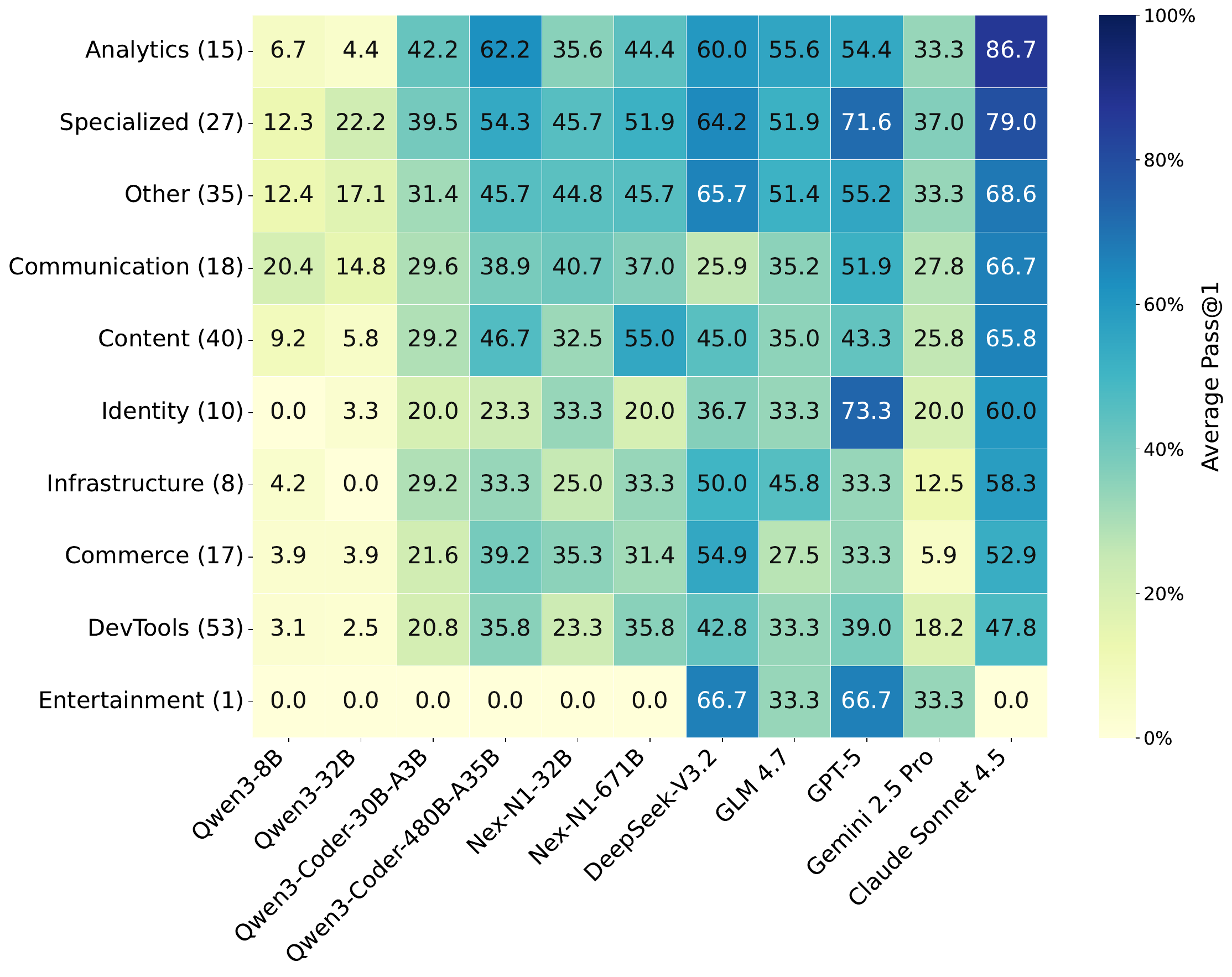}
    \caption{\textbf{Heatmap of Pass@1 accuracy.} The figure compares the performance of various models (x-axis) across different task categories (y-axis). The numbers in parentheses indicate the task count per category, and darker colors represent higher accuracy.}
    \label{fig:category-heatmap}
  \end{minipage}
  \hfill 
  \begin{minipage}[t]{0.48\textwidth}
    \vspace{30pt} 
    \centering
    \includegraphics[width=\linewidth]{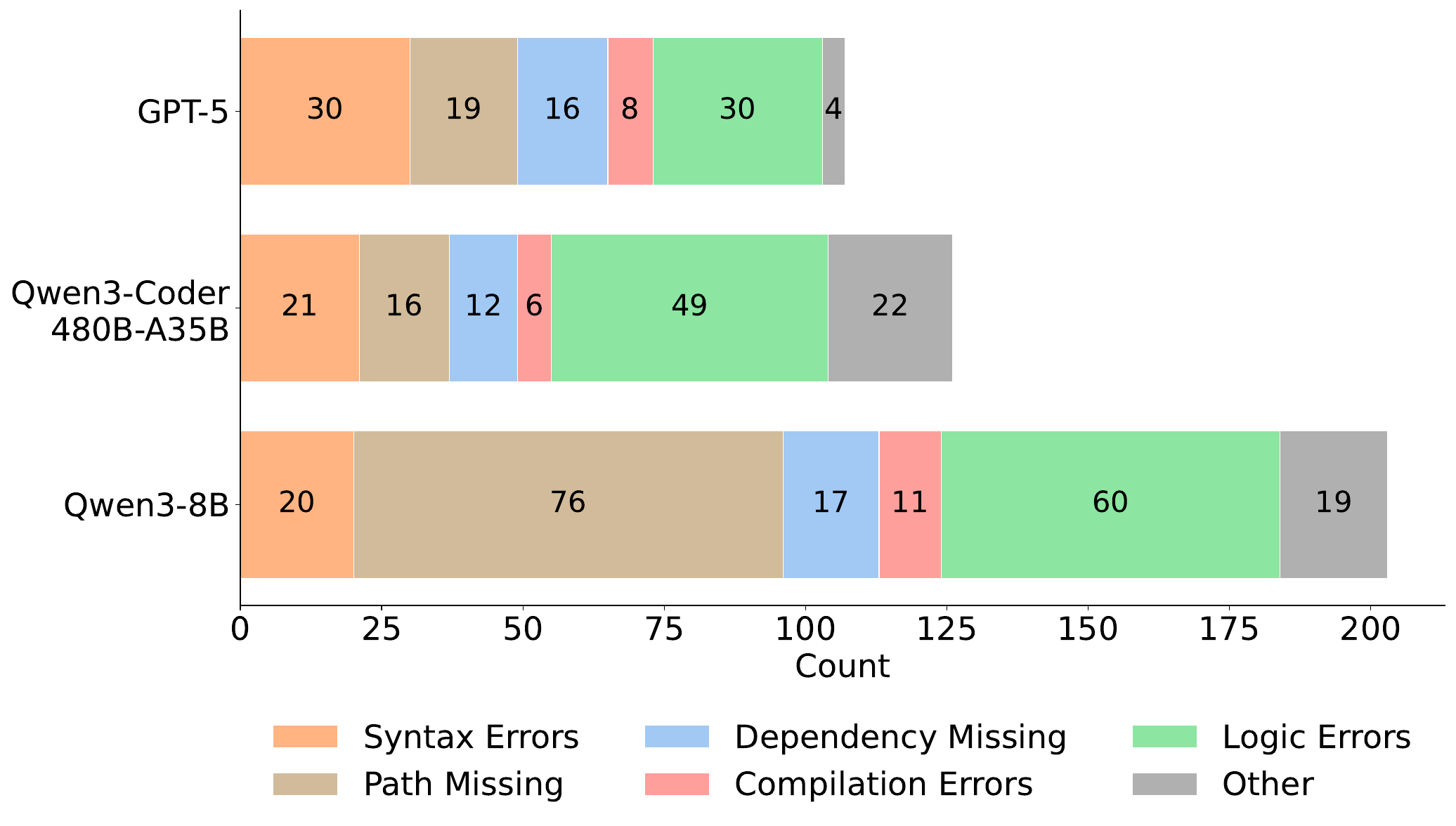}
    \caption{\textbf{Distribution of error types across models.} While environment-related issues remain a persistent challenge for all models, a distinct shift in error composition is observed: smaller models struggle with basic syntax, whereas larger models' failures are concentrated in complex logic.}
    \label{fig:error_type}
  \end{minipage}
\end{figure*}

To better understand the limitations of current code agents, we classify failure cases into six distinct categories, detailed in Table \ref{table:error_type}. Based on this taxonomy, we analyze the error breakdown for three different models, as illustrated in Figure \ref{fig:error_type}. The distribution reveals two critical insights regarding model reliability and the nature of software engineering tasks.

First, environment configuration remains a universal bottleneck, though its severity varies by model capability. Errors related to \textit{Path Missing} and \textit{Dependency Missing} constitute a significant portion of failures. This is particularly acute in the smaller Qwen3-8B model, which suffers disproportionately from fundamental \textit{Path Missing} errors (76 instances)—nearly four times that of GPT-5 (19 instances). This indicates a struggle with basic file system navigation and structure in smaller models.

Second, we observe a correlation between model scale and error sophistication. While the GPT-5 and Qwen3-Coder-480B-A35B-Instruct demonstrate greater robustness in basic syntax and compilation, they exhibit a higher density of \textit{Logic Errors} (30 and 49 instances, respectively) compared to other error types within their own distributions. This shift implies that as models gain better instruction-following and syntactic capabilities, the frontier of failure moves toward high-level reasoning and algorithmic correctness rather than low-level implementation details.

\section{Conclusion}
\label{sec:conclusion}

In this paper, we introduced ABC-Bench, a rigorous evaluation framework designed to assess LLM-based agents' capabilities in handling full-lifecycle backend software engineering tasks. Sourced from 2,000 real-world GitHub repositories, ABC-Bench features a diverse collection of 224 tasks spanning 8 programming languages and 19 distinct frameworks. Our extensive experiments reveal that current systems are still far from reliable in handling these complex tasks. Further analysis identifies environment configuration and deployment as critical bottlenecks, often serving as major barriers to success before code logic can even be validated. These findings underscore a significant gap between current model capabilities and the practical demands of real-world backend engineering. We hope ABC-Bench provides insights that inspire the community to develop agents capable of mastering the full complexity of software production. To support this, we will open-source the code and dataset.












\clearpage
\bibliographystyle{plainnat}
\bibliography{main}

\clearpage
\beginappendix

\startcontents[app]
\begingroup
  \renewcommand{\contentsname}{Appendix Contents}
  \section*{\contentsname}
  \printcontents[app]{}{1}{}
\endgroup
\newpage

\section{Experiment Detail} 
\label{sec:experiment-detail}
To ensure fair and reproducible comparisons across various models and agent frameworks, we standardize our inference and evaluation settings. Specifically, we set the sampling temperature to $1.0$ for reasoning-oriented models and $0.7$ for non-reasoning models. These configurations remain consistent when evaluating different model-framework combinations to maintain a controlled experimental environment.

\textbf{Models.} For closed-source models, as well as DeepSeek-V3.2, Nex-N1, and GLM 4.7, we access them via official APIs, strictly adhering to documented usage policies and recommended request formats. For the Qwen series and the specialized Nex-N1-32B, we deploy them on a cluster of up to 128 NVIDIA H100 GPUs using the SGLang~\citep{zheng2024sglang} framework. To facilitate long-context evaluation for the Qwen3 Instruct family, we extend the context window via YARN-based RoPE scaling during deployment. Furthermore, we enable native tool-calling parsers for Qwen3-8B, Qwen3-32B, Qwen3-Coder, and Nex-N1-32B to assess their tool-use capabilities within their intended inference interfaces.

\textbf{Evaluation.} All evaluations are executed in containerized environments and managed through the Terminal-Bench framework~\citep{tbench_2025}, which orchestrates agents, tasks, and Docker runtime isolation in a uniform way. We evaluate three open-source agent frameworks, OpenHands, mini-SWE-agent, and Claude code. For each framework, we follow the official documentation and widely adopted community configurations. We keep task-level constraints such as timeouts, concurrency, and environment setup consistent across models.

\textbf{Reliability.} To improve reliability and account for stochasticity in decoding and agent behavior, we run the full evaluation for each model three times and report the mean performance across the three runs. Beyond reporting averages, we analyze the error distribution of each run to identify systematic failure patterns and to diagnose run-to-run variability. This procedure helps us maximize the trustworthiness of the execution results and reduces the chance that conclusions are driven by outliers or transient environment effects.

\textbf{Training.} Beyond evaluation, we fine-tune Qwen3 models at 8B and 32B scales using a Megatron-LM-based training stack~\citep{shoeybi2019megatron}. The training is conducted on 128 NVIDIA H100 GPUs, employing a 128K sequence length and context parallelism to ensure efficient scaling. The models are trained for 3 epochs with a target global batch size of 128. Optimization is performed using Adam with a cosine learning rate schedule and a warmup period.

\section{Metric Detail}
\label{sec:metric-detail}

We quantify the agent's capability by decomposing the workflow of the 92 environment-related tasks into two distinct sequential stages: Environment Build ($S_1$) and Functional Execution ($S_2$). This decomposition allows us to calculate the success rate for each stage independently, separating environment configuration outcomes from functional logic correctness.

Stage 1, denoted as Environment Build ($S_1$), involves constructing and starting the service. Success requires generating a valid Dockerfile, building the image, and initializing the container without fatal errors. Stage 2, denoted as Functional Execution ($S_2$), measures the pass rate of functional tests specifically for the subset of tasks that successfully passed $S_1$. It evaluates the correctness of the functional logic within the established environment.

Formally, given a total of $N_{\text{total}}$ environment-related tasks, we track two outcome counts. Let $N_{\text{build}}$ be the number of tasks where the environment is successfully built and started (passing $S_1$), and let $N_{\text{pass}}$ be the number of tasks that pass all functional tests among those that succeeded in $S_1$.

The success rates for these two stages are defined as follows:

\begin{equation}
    \text{Success Rate}(S_1) = \frac{N_{\text{build}}}{N_{\text{total}}}
\end{equation}

\begin{equation}
    \text{Success Rate}(S_2) = \frac{N_{\text{pass}}}{N_{\text{build}}}
\end{equation}

In cases where $N_{\text{build}} = 0$, we define $\text{Success Rate}(S_2) = 0$. This conditional formulation ensures that $S_2$ accurately reflects the model's coding proficiency by decoupling it from the environment configuration bottleneck identified in $S_1$.

\section{Dataset Detail}
\label{sec:dataset-detail}

ABC-Bench is constructed from public GitHub repositories that are explicitly released under the MIT license. We strictly adhere to the terms of these licenses, ensuring that our use is consistent with the open-source nature of the original artifacts. The dataset is intended for research on automated code agents, and we stipulate that any derivatives of this data used for research purposes must remain within research contexts. We list them in Table~\ref{tab:repo_statistics}.

The benchmark focuses on code logic and environment configuration. During the construction pipeline, we operate on repository snapshots and apply automated filtering to detect and redact sensitive credentials, such as access tokens, API keys, and private keys. We explicitly avoid collecting logs or other user-generated content containing personal information. Consequently, the dataset concentrates on technical artifacts rather than personal or demographic data. We do not design tasks to elicit offensive content; any incidental text (e.g., in code comments) is inherited directly from the original open-source projects.

ABC-Bench contains 224 tasks derived from real-world backend repositories, covering a diverse range of programming languages, frameworks, and task categories. Figure~\ref{fig:language-and-framework-full} provides detailed documentation of the dataset statistics, including the distribution over languages, frameworks, and task types. Regarding the data collection protocol, the construction process is primarily automated with internal quality verification by the authors; it does not involve external human participants or manual annotation by crowdworkers, and therefore does not constitute human-subjects research requiring IRB review.

\input{tables/model-detail}

\section{Responsible NLP Research Statements}
\label{sec:ethics-statement}

All code used for task construction in ABC-Pipeline and ABC-Bench is sourced from public GitHub repositories explicitly released under the MIT license. We strictly avoid proprietary code or repositories with unclear licensing. Regarding potential risks, we anticipate minimal negative impact, as our work focuses on evaluating technical infrastructure using public data and does not introduce dangerous capabilities or harmful content.

Publicly-accessible models, datasets, software, and other related artifacts are used under their corresponding licenses or agreements.

Our data collection and task construction processes are primarily automated. To ensure data quality, the authors performed a manual verification of the task titles and descriptions. This review process was conducted exclusively by the research team on a voluntary basis and did not involve the recruitment of external human subjects or crowdworkers. Consequently, no financial compensation was involved. Since this procedure constitutes internal quality assurance rather than experimentation on human subjects, the study is exempt from standard IRB review.

We apply automated screening during dataset construction to detect and exclude content that names or uniquely identifies individual people or contains offensive material. If such cases are detected in the dataset fields we distribute, we remove the affected items or redact sensitive strings to protect privacy and anonymize the data prior to release.

We used AI assistants to support research and writing. All technical claims, experimental results, and reported numbers were produced by our code and verified by the authors.

\input{tables/error_type}
\input{tables/task-category}
\section{Task Instruction}
\label{sec:task-instruction}
The task instructions employed in ABC-Bench are detailed below, derived from the templates in Listing~\ref{list:templete} and Listing~\ref{list:templete-env}. Specifically, Listing~\ref{list:templete} presents the template for tasks without environment configuration, while Listing~\ref{list:templete-env} illustrates the template for tasks that include environment configuration. For a concrete instantiation, please refer to Listing~\ref{list:ABC-Bench}.
\onecolumn

\begin{lstlisting}[
  style=mytext,
  label={list:templete},
  caption={Task template used in ABC-Bench for backend tasks without environment-configuration subtasks.},
  captionpos=t
]
You are a backend development expert. Please inspect the backend project located in the current directory, determine its programming language and architectural style, and then complete the following code implementation and environment setup tasks.

{Task Description}
{API Endpoint Specifications and Requirements}
{Important Notes}
\end{lstlisting}

\begin{lstlisting}[
  style=mytext,
  label={list:templete-env},
  caption={Task template used in ABC-Bench for backend tasks that include environment-configuration subtasks (env subset).},
  captionpos=t
]
You are a backend development expert. Please inspect the backend project located in the current directory, determine its programming language and architectural style, and then complete the following code implementation and environment setup tasks.

{Task Description}
{API Endpoint Specifications and Requirements}
{Important Notes}

After completing all source code implementation, create a Dockerfile for this project using the following example template as a reference (Python version):
```
setup base

FROM nikolaik/python-nodejs:python3.12-nodejs22-bullseye
RUN apt-get update && apt-get install -y sqlite3

install dependencies and copy project files

WORKDIR /app
COPY . /app/
RUN python3 -m pip install -r requirements.txt

ENTRYPOINT ["python3", "app.py"]
```
Notes
1) Ensure all required project dependencies are installed inside the image
2) The generated Dockerfile must successfully build and run the application
3) The Dockerfile must be created in the root directory of the backend project, i.e. {/app/project_name/Dockerfile}
\end{lstlisting}

\begin{lstlisting}[
  style=mytext,
  label={list:ABC-Bench},
  caption={Natural-language task instruction used for ABC-Bench in our environment-centric backend benchmark.},
  captionpos=t
]
You are a backend development expert. Please inspect the backend project located in the current directory, determine its programming language and architectural style, and then complete the following code implementation and environment setup tasks.

Implement the missing comment-domain logic inside:
  module/core/src/main/java/io/zhc1/realworld/service/ArticleCommentService.java

These methods power:
  - POST   /api/articles/{slug}/comments
  - GET    /api/articles/{slug}/comments
  - DELETE /api/articles/{slug}/comments/{id}

Required behavior
1) ArticleComment getComment(int commentId)
   - Look up the comment via ArticleCommentRepository.findById
   - Throw NoSuchElementException with the exact message expected by the REST layer when the id is invalid

2) List<ArticleComment> getComments(Article article)
   - Return every comment associated with the provided article via articleCommentRepository.findByArticle

3) ArticleComment write(ArticleComment articleComment)
   - Persist the comment using the repository and return the managed entity

4) void delete(User requester, ArticleComment articleComment)
   - Reject deletions from non-authors by raising IllegalArgumentException
   - Otherwise remove the entity via articleCommentRepository.delete

Constraints
- Never accept null Article, User, or ArticleComment inputs; throw IllegalArgumentException when misused
- Ownership checks must rely on ArticleComment.isNotAuthor to stay consistent with the aggregate's equality semantics
- Keep the service side-effect free outside repository operations so controllers can safely call it multiple times per request

After completing all source code implementation, create a Dockerfile for this project using the following example template as a reference (Python version):
```
setup base

FROM nikolaik/python-nodejs:python3.12-nodejs22-bullseye
RUN apt-get update && apt-get install -y sqlite3

install dependencies and copy project files

WORKDIR /app
COPY . /app/
RUN python3 -m pip install -r requirements.txt

ENTRYPOINT ["python3", "app.py"]
```
Notes
1) Ensure all required project dependencies are installed inside the image
2) The generated Dockerfile must successfully build and run the application
3) The Dockerfile must be created in the root directory of the backend project, i.e. /app/1chz_realworld-java21-springboot3/Dockerfile
\end{lstlisting}

\input{tables/dataset-detail}
\end{document}

%% file: txt_display.tex
\DeclareFixedFont{\ttb}{T1}{txtt}{bx}{n}{9} 
\DeclareFixedFont{\ttm}{T1}{txtt}{m}{n}{9}  
\definecolor{codedeepblue}{rgb}{0,0,0.5}
\definecolor{codedeepred}{rgb}{0.6,0,0}
\definecolor{codedeepgreen}{rgb}{0,0.5,0}
\definecolor{codegray}{rgb}{0.9,0.9,0.9}
\definecolor{backcolour}{rgb}{0.95,0.95,0.95}

\lstdefinestyle{mytext}{ 
  backgroundcolor=\color{backcolour},  
  basicstyle=\footnotesize\ttfamily,    
     morekeywords={SYSTEM, USER, ASSISTANT, TOOL, SEQUENCE, HIDDEN},
     keywordstyle=\color{black}\bfseries\underbar,
     comment=[s]{```python}{```},
     commentstyle=\color{blue},
  breakatwhitespace=false, 
  breaklines=true,              
  captionpos=b,                 
  frame=tb,	                   	
  keepspaces=true,                
  rulecolor=\color{white},        
  showspaces=false,                
  showstringspaces=false,          
  showtabs=false,                  
  stepnumber=1,                   
  tabsize=2
}

%% file: tables/compare.tex
\renewcommand{\arraystretch}{1} 

\begin{wraptable}{r}{0.5\textwidth}
    \centering
    \vspace{-10pt} 
    
    \setlength{\tabcolsep}{2.5pt}
    \small 
    
    \resizebox{\linewidth}{!}{%
        \begin{tabular}{lccccc}
            \toprule
            \textbf{Benchmark} & 
            \textbf{Expl.} & 
            \textbf{Code} & 
            \textbf{Env.} & 
            \textbf{Deploy} & 
            \textbf{E2E} \\
            \midrule
            BaxBench~\cite{vero2025baxbench}        &        & \cmark &        &        & \cmark \\
            SWE-bench~\cite{jin2024swebench}       & \cmark & \cmark &        &        &        \\
            FullStack Bench~\cite{liu2024fullstackbenchevaluatingllms} & \cmark & \cmark &        &        &        \\
            DevBench~\cite{li2024devbench}        &        & \cmark & \cmark &        &        \\
            \midrule
            \rowcolor{lightgreen} 
            \textbf{ABC-Bench (Ours)} & \cmark & \cmark & \cmark & \cmark & \cmark \\
            \bottomrule
        \end{tabular}%
    }
    \vspace{1mm}
    
    \caption{
        \textbf{Comparison of Benchmark Scope.} 
        Real-world backend development mandates a continuous workflow spanning five distinct stages: 
        (1) \textit{Repository Exploration (Expl.)}, 
        (2) \textit{Code Editing (Code)}, 
        (3) \textit{Environment Setup (Env.)}, 
        (4) \textit{Deployment (Deploy)}, and 
        (5) \textit{End-to-End Testing (E2E)}. 
        While prior benchmarks focus primarily on localized coding tasks, ABC-Bench is the only benchmark to encompass this entire lifecycle.
    }
    \label{tab:benchmark-coverage}
    
\end{wraptable}

%% file: tables/main-results.tex

\begin{table*}[t]
\belowrulesep=0pt
\aboverulesep=0pt
\fontsize{10.5}{14}\selectfont
\centering

\resizebox{\textwidth}{!}{
\begin{tabular}{l c *{8}{c} c}
\toprule[1.5pt]
\multirow{2}{*}{\textbf{Model}} &
\multirow{2}{*}{\textbf{Thinking}} &
\multicolumn{8}{c}{\textbf{By language (Avg.\ pass@1, \%)}} &
\multirow{2}{*}{\textbf{Overall}} \\
\cmidrule(lr){3-10}
 &  & Py & Go & JS & Java & Ruby & C\# & PHP & Rust & \\
\midrule
\rowcolor{gray!15}
\multicolumn{11}{c}{\textit{Open Source Models}} \\
\addlinespace[0.2em]
Qwen3-8B               & Yes & 26.7          &  5.7          & 21.9          &  3.9          &  7.1          &  4.7          &  1.0          &  0.0          &  8.3$\pm$1.1 \\
Qwen3-32B              & Yes & 16.7          &  2.3          & 22.8          &  2.9          &  9.1          &  7.8          &  5.1          &  0.0          &  8.9$\pm$1.1 \\
Qwen3-Coder-30B-A3B    & No  & 46.7          & 31.0          & \uline{50.9}          & 19.6          & 25.3          & 13.2          & 31.3          &  0.0          & 28.6$\pm$1.7 \\
Qwen3-Coder-480B-A35B  & No  & \uline{56.7}          & 55.2          & \textbf{57.0} & 52.9          & \uline{36.4}          & 24.0          & \uline{36.4}          & \textbf{16.7} & \uline{43.1$\pm$1.9} \\
Nex-N1-32B             & No  & 43.3          & \textbf{67.8} & 49.1          & 26.5          & 22.2          & 20.9          & 28.3          &  0.0          & 34.5$\pm$1.8 \\
Nex-N1-671B            & No  & \textbf{60.0} & 63.2          & 49.1          & \uline{54.9}          & 30.3          & 26.4          & 34.3          &  0.0          & 42.1$\pm$1.9 \\
DeepSeek-V3.2   & No  & \uline{56.7}          & \uline{64.4}          & \textbf{57.0} & \textbf{57.8} &\textbf{ 42.4} & \textbf{38.8} & \textbf{48.5} &  0.0          & \textbf{50.1$\pm$1.9} \\
GLM 4.7                & Yes  & 40.0          & 56.3          & 50.0          & 40.2          & 34.3          & \uline{32.6}          & \uline{36.4}          &  0.0          & 40.1$\pm$1.9 \\
\midrule
\rowcolor{gray!15}
\multicolumn{11}{c}{\textit{Proprietary Models}} \\
\addlinespace[0.2em]
GPT-5                  & Yes & \uline{30.0}          & \uline{56.9}          & 46.5          & \textbf{67.6} & \textbf{55.6} & \uline{36.4}          & \uline{44.4}          & \textbf{41.7} & \uline{49.4$\pm$1.9} \\
Gemini 2.5 Pro         & Yes & \uline{30.0}          & 29.9          & \uline{47.4}          & \uline{19.6}          & 20.2          & 17.1          & 16.2          &  8.3          & 25.0$\pm$1.7 \\
Claude Sonnet 4.5      & Yes & \textbf{73.3} & \textbf{75.9} & \textbf{69.3} & \textbf{67.6} & \uline{51.5}          & \textbf{46.5} & \textbf{74.7} & \uline{33.3}          & \textbf{63.2$\pm$1.9} \\
\bottomrule[1.5pt]
\end{tabular}}

\caption{
ABC-Bench results measured by average pass@1 (\%) under three independent attempts per task.
We report the overall performance over all 224 tasks, along with a breakdown by programming language. Models are evaluated using OpenHands as the agent framework. The highest scores for open-source and proprietary models are highlighted in bold, respectively; the second-highest scores are underlined, except for values equal to 0.
}
\label{table:main-results}
\end{table*}

%% file: tables/model-detail.tex
\begin{table*}[t]
\centering
\setlength{\tabcolsep}{3.5pt}
\renewcommand{\arraystretch}{1.15}

\begin{tabularx}{\textwidth}{l c c c c c c}
\toprule
\textbf{Model} & \textbf{Params} & \textbf{Launch} & \textbf{Max tokens} & \textbf{MLP Type} & \textbf{Provider} & \textbf{License} \\
\midrule
Qwen3-8B & 8B & 2025--04 & 128K & Dense & Alibaba & Apache 2.0 \\
Qwen3-32B & 32B & 2025--04 & 128K & Dense & Alibaba & Apache 2.0 \\
Qwen3-Coder-30B-A3B & 30B & 2025--07 & 256K & MoE & Alibaba & Apache 2.0 \\
Qwen3-Coder-480B-A35B & 480B & 2025--07 & 256K & MoE & Alibaba & Apache 2.0 \\
Nex-N1-32B & 32B & 2025--11 & 128K & Dense & Nex AGI & Apache 2.0 \\
Nex-N1-671B & 671B & 2025--11 & 128K & MoE & Nex AGI & Apache 2.0 \\
DeepSeek-V3.2 & 671B & 2025--12 & 128K & MoE & DeepSeek & MIT \\
GLM 4.7 & 358B & 2025--12 & 200K & MoE & Z.ai & MIT \\
GPT-5 & -- & 2025--08 & 400K & -- & OpenAI & Proprietary \\
Gemini 2.5 Pro & -- & 2025--06 & 1M & -- & Google DM & Proprietary \\
Claude Sonnet 4.5 & -- & 2025--09 & 200K & -- & Anthropic & Proprietary \\
\bottomrule
\end{tabularx}

\caption{Model configurations used in our experiments, including parameter scale, launch date, maximum context length, MLP type, provider, and license.}
\label{table:model-detail}
\end{table*}

%% file: tables/error_type.tex
\begin{table*}[t]
  \centering
  \begin{tabular}{p{0.20\textwidth} p{0.75\textwidth}}
    \toprule
    \textbf{Error type} & \textbf{Meaning} \\
    \midrule

    Syntax Errors
    & Static syntax issues in source code or configuration files, reported by a
      compiler, interpreter, or linter. Typical examples include invalid tokens,
      missing brackets, malformed JSON or YAML, or other parse errors that prevent
      the code from being compiled or executed. \vspace{0.6em} \\

    Path Missing
    & A required file or directory cannot be found when building the image or
      running the service. Common cases include incorrect paths in Dockerfile
      \texttt{COPY} or \texttt{ADD} instructions, missing project directories or
      entrypoints, or startup scripts referencing non-existent configuration
      files. \vspace{0.6em} \\

    Dependency Missing
    & A required dependency is not available in the environment. This includes
      missing language packages, system libraries, or runtime extensions, such as
      missing PHP extensions, Python or Node.js packages, Java runtimes, or
      operating-system-level libraries that the application or framework expects.
      \vspace{0.6em} \\

    Compilation Errors
    & Build or compilation failures that are not purely syntax mistakes, but arise
      during the compilation or linking phase. Typical examples are type
      mismatches, unresolved symbols, missing headers, or incompatibilities
      between language, framework, or tool versions that prevent producing a
      build artifact. \vspace{0.6em} \\

    Logic Errors
    & The service builds and runs, but the behavior is functionally incorrect.
      This includes wrong business logic, incorrect HTTP status codes, mismatched
      response schemas or data types, failing assertions in tests, or incorrect
      handling of edge cases, even though the process itself does not necessarily
      crash. \vspace{0.6em} \\

    Other
    & Errors that do not fit cleanly into the above categories, typically related
      to environment, infrastructure, or multi-factor issues. Examples include
      port conflicts, file permission problems, disk or memory limits, SSL or
      network connectivity failures, problems with base images, or mixed cases
      where no single category is clearly dominant. \\
    \bottomrule
  \end{tabular}

  \caption{Explanations of error types used in our analysis, summarizing common failure modes observed during code changes, build, deployment, and end-to-end API evaluation.}
  \label{table:error_type}
\end{table*}

%% file: tables/task-category.tex
\begin{table*}[t]
\centering
\renewcommand{\arraystretch}{1.0}

\begin{tabular}{p{2.8cm} c p{10cm}}
\toprule
\textbf{Category} & \textbf{Count} & \textbf{Description} \\
\midrule
Content & 34 & Content platforms such as blogs, knowledge bases, documentation sites, and course portals. \vspace{0.6em}\\
Commerce & 17 & Backends for e-commerce, ordering, payment processing, bookkeeping, and financial workflows. \vspace{0.6em}\\
DevTools & 41 & SDKs, frameworks, scaffolds, routers, and package tools targeting developers. \vspace{0.6em}\\
Analytics & 14 & Services focused on search, reporting, monitoring, recommendations, and other data analytics. \vspace{0.6em}\\
Communication & 18 & Chat, email, notification, collaborative editing, and CRM-style systems. \vspace{0.6em}\\
Entertainment & 1 & Backends serving games or other entertainment experiences. \vspace{0.6em}\\
Identity & 9 & Systems handling login, OAuth, SSO, and authorization management. \vspace{0.6em}\\
Infrastructure & 2 & Examples centered on containerization, operations, and deployment practices. \vspace{0.6em}\\
Specialized & 36 & REST/GraphQL services or demos built around specific business domains. \vspace{0.6em}\\
Other & 33 & Mixed scenarios. \vspace{0.6em}\\
\bottomrule
\end{tabular}

\caption{Task categories in ABC-Bench, grouped by backend application domain with counts and brief descriptions.}
\label{table:task-category}
\end{table*}

%% file: tables/dataset-detail.tex
\clearpage
\begingroup
\small
\renewcommand{\arraystretch}{0.985} 
\onecolumn
\begin{longtable}[c]{l | l p{0.65\textwidth} c}
\caption{Repository statistics: ABC-Bench comprises a diverse collection of 127 GitHub repositories and 224 tasks.}
\label{tab:repo_statistics} \\
\toprule
\textbf{Language} & \textbf{Framework} & \textbf{Repo} & \textbf{\# Tasks} \\\midrule
\endfirsthead

\caption[]{Repository statistics (continued)} \\
\toprule
\textbf{Language} & \textbf{Framework} & \textbf{Repo} & \textbf{\# Tasks} \\\midrule
\endhead

\midrule
\multicolumn{4}{r}{\textit{Continued on next page}} \\
\endfoot

\bottomrule
\endlastfoot

      \multirow{33}{*}{C\#}                    & \multirow{33}{*}{ASP.NET Core} & \href{https://github.com/angelsix/fasetto-word}{angelsix/fasetto-word} &  4 \\
                                               &                                & \href{https://github.com/AIDotNet/OpenDeepWiki}{AIDotNet/OpenDeepWiki} &  3 \\
                                               &                                & \href{https://github.com/neozhu/cleanaspire}{neozhu/cleanaspire} &  3 \\
                                               &                                & \href{https://github.com/cloudinary/CloudinaryDotNet}{cloudinary/CloudinaryDotNet} &  2 \\
                                               &                                & \href{https://github.com/RageAgainstThePixel/OpenAI-DotNet}{RageAgainstThePixel/OpenAI-DotNet} &  2 \\
                                               &                                & \href{https://github.com/BenediktAlkin/SongTaggerForSpotify}{BenediktAlkin/SongTaggerForSpotify} &  2 \\
                                               &                                & \href{https://github.com/ntxinh/AspNetCore-DDD}{ntxinh/AspNetCore-DDD} &  1 \\
                                               &                                & \href{https://github.com/withsalt/BilibiliLiveTools}{withsalt/BilibiliLiveTools} &  1 \\
                                               &                                & \href{https://github.com/BlazorStatic/BlazorStatic}{BlazorStatic/BlazorStatic} &  1 \\
                                               &                                & \href{https://github.com/ardalis/CachedRepository}{ardalis/CachedRepository} &  1 \\
                                               &                                & \href{https://github.com/ardalis/CleanArchitecture}{ardalis/CleanArchitecture} &  1 \\
                                               &                                & \href{https://github.com/stidsborg/Cleipnir.NET}{stidsborg/Cleipnir.NET} &  1 \\
                                               &                                & \href{https://github.com/linezero/GitServer}{linezero/GitServer} &  1 \\
                                               &                                & \href{https://github.com/khellang/Middleware}{khellang/Middleware} &  1 \\
                                               &                                & \href{https://github.com/paiden/Nett}{paiden/Nett} &  1 \\
                                               &                                & \href{https://github.com/devmentors/PackIT}{devmentors/PackIT} &  1 \\
                                               &                                & \href{https://github.com/reactjs/React.NET}{reactjs/React.NET} &  1 \\
                                               &                                & \href{https://github.com/samanazadi1996/Sam.CleanArchitecture}{samanazadi1996/Sam.CleanArchitecture} &  1 \\
                                               &                                & \href{https://github.com/sharpenrocks/Sharpen}{sharpenrocks/Sharpen} &  1 \\
                                               &                                & \href{https://github.com/Havunen/SystemTextJsonPatch}{Havunen/SystemTextJsonPatch} &  1 \\
                                               &                                & \href{https://github.com/microsoftgraph/aspnetcore-webhooks-sample}{microsoftgraph/aspnetcore-webhooks-sample} &  1 \\
                                               &                                & \href{https://github.com/dtm-labs/client-csharp}{dtm-labs/client-csharp} &  1 \\
                                               &                                & \href{https://github.com/jamesmh/coravel}{jamesmh/coravel} &  1 \\
                                               &                                & \href{https://github.com/Azure-Samples/cosmos-db-design-patterns}{Azure-Samples/cosmos-db-design-patterns} &  1 \\
                                               &                                & \href{https://github.com/Azure/dev-spaces}{Azure/dev-spaces} &  1 \\
                                               &                                & \href{https://github.com/dotnet/dotNext}{dotnet/dotNext} &  1 \\
                                               &                                & \href{https://github.com/cornflourblue/dotnet-6-crud-api}{cornflourblue/dotnet-6-crud-api} &  1 \\
                                               &                                & \href{https://github.com/cornflourblue/dotnet-6-jwt-refresh-tokens-api}{cornflourblue/dotnet-6-jwt-refresh-tokens-api} &  1 \\
                                               &                                & \href{https://github.com/dotnet/dotnet-monitor}{dotnet/dotnet-monitor} &  1 \\
                                               &                                & \href{https://github.com/Azure-Samples/eShopOnAzure}{Azure-Samples/eShopOnAzure} &  1 \\
                                               &                                & \href{https://github.com/unosquare/passcore}{unosquare/passcore} &  1 \\
                                               &                                & \href{https://github.com/ThunderDev1/reactjs-ts-identityserver}{ThunderDev1/reactjs-ts-identityserver} &  1 \\
                                               &                                & \href{https://github.com/simplyvinay/vue-expenses}{simplyvinay/vue-expenses} &  1 \\
      \midrule  
      \multirow{22}{*}{JavaScript}             & \multirow{21}{*}{Express}      & \href{https://github.com/15Dkatz/official_joke_api}{15Dkatz/official\_joke\_api} &  4 \\
                                               &                                & \href{https://github.com/CapacitorSet/box-js}{CapacitorSet/box-js} &  3 \\
                                               &                                & \href{https://github.com/azat-co/expressworks}{azat-co/expressworks} &  3 \\
                                               &                                & \href{https://github.com/auth0/auth0-react}{auth0/auth0-react} &  2 \\
                                               &                                & \href{https://github.com/carteb/carte-blanche}{carteb/carte-blanche} &  2 \\
                                               &                                & \href{https://github.com/conclave-team/conclave}{conclave-team/conclave} &  2 \\
                                               &                                & \href{https://github.com/jellyfangs/messenger-bot-tutorial}{jellyfangs/messenger-bot-tutorial} &  2 \\
                                               &                                & \href{https://github.com/BretFisher/node-docker-good-defaults}{BretFisher/node-docker-good-defaults} &  2 \\
                                               &                                & \href{https://github.com/hagopj13/node-express-boilerplate}{hagopj13/node-express-boilerplate} &  2 \\
                                               &                                & \href{https://github.com/eduardoboucas/staticman}{eduardoboucas/staticman} &  2 \\
                                               &                                & \href{https://github.com/filipedeschamps/video-maker}{filipedeschamps/video-maker} &  2 \\
                                               &                                & \href{https://github.com/adrianvlupu/C4-Builder}{adrianvlupu/C4-Builder} &  1 \\
                                               &                                & \href{https://github.com/microsoft/PowerBI-Developer-Samples}{microsoft/PowerBI-Developer-Samples} &  1 \\
                                               &                                & \href{https://github.com/davila7/claude-code-templates}{davila7/claude-code-templates} &  1 \\
                                               &                                & \href{https://github.com/attacomsian/code-examples}{attacomsian/code-examples} &  1 \\
                                               &                                & \href{https://github.com/external-secrets/kubernetes-external-secrets}{external-secrets/kubernetes-external-secrets} &  1 \\
                                               &                                & \href{https://github.com/jagenjo/litegraph.js}{jagenjo/litegraph.js} &  1 \\
                                               &                                & \href{https://github.com/kogosoftwarellc/open-api}{kogosoftwarellc/open-api} &  1 \\
                                               &                                & \href{https://github.com/ankur-anand/simple-sso}{ankur-anand/simple-sso} &  1 \\
                                               &                                & \href{https://github.com/daspinola/video-stream-sample}{daspinola/video-stream-sample} &  1 \\
                                               &                                & \href{https://github.com/welldone-software/why-did-you-render}{welldone-software/why-did-you-render} &  1 \\
      \cmidrule{2-4}  
                                               & \multirow{1}{*}{Koa}           & \href{https://github.com/jamhall/s3rver}{jamhall/s3rver} &  2 \\
      \midrule  
      \multirow{21}{*}{Java}                   & \multirow{19}{*}{Spring Boot}  & \href{https://github.com/1chz/realworld-java21-springboot3}{1chz/realworld-java21-springboot3} &  6 \\
                                               &                                & \href{https://github.com/getmoneynote/moneynote-api}{getmoneynote/moneynote-api} &  3 \\
                                               &                                & \href{https://github.com/kawhii/sso}{kawhii/sso} &  3 \\
                                               &                                & \href{https://github.com/dailycodebuffer/Spring-MVC-Tutorials}{dailycodebuffer/Spring-MVC-Tutorials} &  2 \\
                                               &                                & \href{https://github.com/ttulka/ddd-example-ecommerce}{ttulka/ddd-example-ecommerce} &  2 \\
                                               &                                & \href{https://github.com/cassiomolin/log-aggregation-spring-boot-elastic-stack}{cassiomolin/log-aggregation-spring-boot-elastic-stack} &  2 \\
                                               &                                & \href{https://github.com/AwakenCN/Almost-Famous}{AwakenCN/Almost-Famous} &  1 \\
                                               &                                & \href{https://github.com/zzzzbw/Fame}{zzzzbw/Fame} &  1 \\
                                               &                                & \href{https://github.com/KonBAI-Q/RuoYi-Flowable-Plus}{KonBAI-Q/RuoYi-Flowable-Plus} &  1 \\
                                               &                                & \href{https://github.com/HouariZegai/Tutorials}{HouariZegai/Tutorials} &  1 \\
                                               &                                & \href{https://github.com/liuxx-u/bird-java}{liuxx-u/bird-java} &  1 \\
                                               &                                & \href{https://github.com/cucumber/cucumber-jvm}{cucumber/cucumber-jvm} &  1 \\
                                               &                                & \href{https://github.com/PomZWJ/database-export}{PomZWJ/database-export} &  1 \\
                                               &                                & \href{https://github.com/liyupi/father}{liyupi/father} &  1 \\
                                               &                                & \href{https://github.com/zhongjinggz/geekdemo}{zhongjinggz/geekdemo} &  1 \\
                                               &                                & \href{https://github.com/threedr3am/learnjavabug}{threedr3am/learnjavabug} &  1 \\
                                               &                                & \href{https://github.com/ssssssss-team/magic-api}{ssssssss-team/magic-api} &  1 \\
                                               &                                & \href{https://github.com/nonacosa/new-bee}{nonacosa/new-bee} &  1 \\
                                               &                                & \href{https://github.com/allanzhuo/yyblog}{allanzhuo/yyblog} &  1 \\
      \cmidrule{2-4}  
                                               & \multirow{2}{*}{Spring MVC}    & \href{https://github.com/3pillarlabs/socialauth}{3pillarlabs/socialauth} &  2 \\
                                               &                                & \href{https://github.com/v5tech/elasticsearch-jest-example}{v5tech/elasticsearch-jest-example} &  1 \\
      \midrule  
      \multirow{16}{*}{Ruby}                   & \multirow{8}{*}{Ruby on Rails} & \href{https://github.com/danielschuster-muc/potter-db}{danielschuster-muc/potter-db} &  6 \\
                                               &                                & \href{https://github.com/exoego/rspec-openapi}{exoego/rspec-openapi} &  6 \\
                                               &                                & \href{https://github.com/active-elastic-job/active-elastic-job}{active-elastic-job/active-elastic-job} &  3 \\
                                               &                                & \href{https://github.com/amatsuda/html5_validators}{amatsuda/html5\_validators} &  2 \\
                                               &                                & \href{https://github.com/rubygems/rubygems.org}{rubygems/rubygems.org} &  2 \\
                                               &                                & \href{https://github.com/hyperstack-org/hyperstack}{hyperstack-org/hyperstack} &  1 \\
                                               &                                & \href{https://github.com/ElMassimo/oj_serializers}{ElMassimo/oj\_serializers} &  1 \\
                                               &                                & \href{https://github.com/crmne/ruby_llm}{crmne/ruby\_llm} &  1 \\
      \cmidrule{2-4}  
                                               & \multirow{6}{*}{Sinatra}       & \href{https://github.com/assaf/rack-oauth2-server}{assaf/rack-oauth2-server} &  2 \\
                                               &                                & \href{https://github.com/maccman/go}{maccman/go} &  1 \\
                                               &                                & \href{https://github.com/mobomo/green_onion}{mobomo/green\_onion} &  1 \\
                                               &                                & \href{https://github.com/isaiah/jubilee}{isaiah/jubilee} &  1 \\
                                               &                                & \href{https://github.com/brandur/rocket-rides-atomic}{brandur/rocket-rides-atomic} &  1 \\
                                               &                                & \href{https://github.com/joakimk/testbot}{joakimk/testbot} &  1 \\
      \cmidrule{2-4}  
                                               & \multirow{2}{*}{Rack}          & \href{https://github.com/flippercloud/flipper}{flippercloud/flipper} &  3 \\
                                               &                                & \href{https://github.com/amoniacou/danthes}{amoniacou/danthes} &  1 \\
      \midrule  
      \multirow{14}{*}{PHP}                    & \multirow{8}{*}{Laravel}       & \href{https://github.com/amsgames/laravel-shop}{amsgames/laravel-shop} &  5 \\
                                               &                                & \href{https://github.com/cretueusebiu/laravel-vue-spa}{cretueusebiu/laravel-vue-spa} &  5 \\
                                               &                                & \href{https://github.com/laqul/laqul}{laqul/laqul} &  3 \\
                                               &                                & \href{https://github.com/JeffreyWay/council}{JeffreyWay/council} &  2 \\
                                               &                                & \href{https://github.com/ploi/roadmap}{ploi/roadmap} &  2 \\
                                               &                                & \href{https://github.com/WangNingkai/OLAINDEX}{WangNingkai/OLAINDEX} &  1 \\
                                               &                                & \href{https://github.com/devinsays/laravel-react-bootstrap}{devinsays/laravel-react-bootstrap} &  1 \\
                                               &                                & \href{https://github.com/protonemedia/laravel-splade}{protonemedia/laravel-splade} &  1 \\
      \cmidrule{2-4}  
                                               & \multirow{5}{*}{Symfony}       & \href{https://github.com/composer/packagist}{composer/packagist} &  4 \\
                                               &                                & \href{https://github.com/AdventureLookup/AdventureLookup}{AdventureLookup/AdventureLookup} &  2 \\
                                               &                                & \href{https://github.com/hgraca/explicit-architecture-php}{hgraca/explicit-architecture-php} &  2 \\
                                               &                                & \href{https://github.com/nelmio/NelmioJsLoggerBundle}{nelmio/NelmioJsLoggerBundle} &  1 \\
                                               &                                & \href{https://github.com/ankitpokhrel/tus-php}{ankitpokhrel/tus-php} &  1 \\
      \cmidrule{2-4}  
                                               & \multirow{1}{*}{Silex}         & \href{https://github.com/intaro/pinboard}{intaro/pinboard} &  3 \\
      \midrule  
      \multirow{14}{*}{Go}                     & \multirow{7}{*}{net/http}      & \href{https://github.com/tsileo/blobstash}{tsileo/blobstash} &  6 \\
                                               &                                & \href{https://github.com/go-spatial/tegola}{go-spatial/tegola} &  2 \\
                                               &                                & \href{https://github.com/flycash/toy-web}{flycash/toy-web} &  2 \\
                                               &                                & \href{https://github.com/Azure/azure-sdk-for-go}{Azure/azure-sdk-for-go} &  1 \\
                                               &                                & \href{https://github.com/99designs/gqlgen}{99designs/gqlgen} &  1 \\
                                               &                                & \href{https://github.com/darkweak/souin}{darkweak/souin} &  1 \\
                                               &                                & \href{https://github.com/stripe-archive/timberlake}{stripe-archive/timberlake} &  1 \\
      \cmidrule{2-4}  
                                               & \multirow{5}{*}{Gin}           & \href{https://github.com/air-go/rpc}{air-go/rpc} &  3 \\
                                               &                                & \href{https://github.com/Email-Dashboard/Email-Dashboard}{Email-Dashboard/Email-Dashboard} &  2 \\
                                               &                                & \href{https://github.com/izghua/go-blog}{izghua/go-blog} &  1 \\
                                               &                                & \href{https://github.com/nhost/hasura-auth}{nhost/hasura-auth} &  1 \\
                                               &                                & \href{https://github.com/swaggo/swag}{swaggo/swag} &  1 \\
      \cmidrule{2-4}  
                                               & \multirow{1}{*}{Chi}           & \href{https://github.com/ashirt-ops/ashirt-server}{ashirt-ops/ashirt-server} &  4 \\
      \cmidrule{2-4}  
                                               & \multirow{1}{*}{CleverGo}      & \href{https://github.com/clevergo/clevergo}{clevergo/clevergo} &  3 \\
      \midrule  
      \multirow{5}{*}{Python}                  & \multirow{2}{*}{Flask}         & \href{https://github.com/HaoZhang95/Python24}{HaoZhang95/Python24} &  3 \\
                                               &                                & \href{https://github.com/stripe-samples/accept-a-payment}{stripe-samples/accept-a-payment} &  2 \\
      \cmidrule{2-4}  
                                               & \multirow{2}{*}{FastAPI}       & \href{https://github.com/Azure/apiops}{Azure/apiops} &  3 \\
                                               &                                & \href{https://github.com/Tongjilibo/bert4torch}{Tongjilibo/bert4torch} &  1 \\
      \cmidrule{2-4}  
                                               & \multirow{1}{*}{AWS SAM}       & \href{https://github.com/aws-samples/serverless-test-samples}{aws-samples/serverless-test-samples} &  1 \\
      \midrule  
      \multirow{2}{*}{Rust}                    & \multirow{2}{*}{Axum}          & \href{https://github.com/restsend/rustpbx}{restsend/rustpbx} &  3 \\
                                               &                                & \href{https://github.com/Totodore/socketioxide}{Totodore/socketioxide} &  1 \\

\end{longtable}
\endgroup